# A Deep Learning Approach for the Detection of COVID-19 from Chest X-Ray images using Convolutional Neural Networks

Aditya Saxena*, *Birla Institute of Technology and Science, Pilani, Dubai Campus*
E-mail: *f20190089@dubai.bits-pilani.ac.in*

Shamsheer Pal Singh, *Birla Institute of Technology and Science, Pilani, Dubai Campus*
E-mail: *f20190055@dubai.bits-pilani.ac.in*

Vilas H Gaidhane, *Birla Institute of Technology and Science, Pilani, Dubai Campus*
*Academic City, Dubai, UAE*

**Abstract**

The COVID-19 (coronavirus) is an ongoing pandemic caused by severe acute respiratory syndrome coronavirus 2 (SARS-CoV-2). The virus was first identified in mid-December 2019 in the Hubei province of Wuhan, China and by now has spread throughout the planet with more than 75.5 million confirmed cases and more than 1.67 million deaths. With limited number of COVID-19 test kits available in medical facilities, it is important to develop and implement an automatic detection system as an alternative diagnosis option for COVID-19 detection that can used on a commercial scale. Chest X-ray is the first imaging technique that plays an important role in the diagnosis of COVID-19 disease. Computer vision and deep learning techniques can help in determining COVID-19 virus with Chest X-ray Images. Due to the high availability of large-scale annotated image datasets, great success has been achieved using convolutional neural network for image analysis and classification. In this research, we have proposed a deep convolutional neural network trained on five open access datasets with binary output: Normal and Covid. The performance of the model is compared with four pre-trained convolutional neural network-based models (COVID-Net, ResNet18, ResNet and MobileNet-V2) and it has been seen that the proposed model provides better accuracy on the validation set as compared to the other four pre-trained models. This research work provides promising results which can be further improvise and implement on a commercial scale.

## 1. INTRODUCTION

The coronavirus (COVID-19) pandemic has affected billions of people since the time of its emergence from Wuhan, China in December 2019.[1] The virus led to an outbreak at a very fast rate. A lot of research was conducted to identify the type of virus that caused COVID-19 disease and it was concluded that it belonged to a huge family of respiratory viruses that can cause diseases such as Middle East Respiratory Syndrome (MERS-CoV) and Severe Acute Respiratory Syndrome (SARS-CoV). The new SARS-CoV-2 virus can develop viral pneumonia. The population has witnessed a very high mortality rate in some states. The death toll around the world is increasing day by day. Therefore, it is necessary to develop an accurate, fast and cost-effective tool for diagnosis of viral pneumonia. This will serve as the initial step for taking further preventive measures like isolation, contact tracing and treatment for stopping the outbreak.

One popular method to detect the virus is viral nucleic acid detection using real-time polymerase chain reaction, also known as RT-PCR test.[2] This test is very sensitive and has several limitations. For example, it cannot detect coronavirus developed before taking DNA sequence samples. Moreover, it takes 2-3 days to produce the result and requires many arrangements, public space. Many countries are not able to provide these conditions for testing of thousands of patients. Hence, continuing this method might slow down the process of controlling the pandemic.[3]

In this scenario, medical imaging can prove to be a vital technique for diagnosis. Chest radiography plays an important role in the early diagnosis of pneumonia. It is commonly used because of its fast-imaging speed and low cost.[4] However accurate and fast diagnosis of a X-Ray image is only possible with the help of expert knowledge.[5][6]

The common diagnosis is done based on pneumonia symptoms (fever, chills, dry cough) but due to many asymptomatic patients being tested positive, it is necessary to improve the screening process by taking the help of X-ray images and testing more people as soon as possible. Due to increasing cases and a smaller number of specialists available to make diagnosis, the screening process becomes a tough task. Hence, doctors must depend on machine learning models for a fast and accurate diagnosis. Several machine learning approaches have already been used for analysing the X-ray images.[7]

Traditional methods like support vector methods (SVMs) have several disadvantages. Over the years, their performance has degraded and is not considered at par with practical standards. Moreover, their development is very time-consuming.

Deep learning approaches have led to major advancements in the field of medical image classification and has become an effective tool for doctors to analyse the images and diagnose the problem. The breakthroughs have made them capable of carrying out many existing medical image analysis tasks like detection, staging and description of pathological abnormalities. Convolutional Neural Network (CNN) is one popular approach for analysing images, and it has made remarkable achievements in the medical field.[8]

Deep Convolutional Networks (DCNNs) are being constructed to analyse chest images and diagnose common thorax diseases and differentiate between viral pneumonia and non-viral pneumonia.[9][10] While many common viruses like influenza A/B, chickenpox, coronaviruses, and measles can cause pneumonia, the ones with viral pneumonia cause substantial differences in X-Ray images. Which means that every case of viral pneumonia will contain variable visual appearances. Moreover, finding a dataset with positive samples poses another problem. Therefore, it is crucial to develop a model which can overcome these pathological abnormalities and detect the virus with high accuracy.

These methods are being used in the medical field since 2012 and have shown significantly better performance than other methods. CheXNet, a CNN with 121 layers which was trained on ChestX-ray 14 dataset having 112,120 images of frontal-view chest X-rays performed better than the average performance of four radiologists [11]

CNN has the ability to learn automatically from domain-specific images and hence differentiates itself from classical machine learning methods. Different strategies can be implemented to train CNN architecture to acquire the desired accuracy and results. In this paper, we have used a similar model of deep convolutional neural network for the analysis of chest X-Rays. The collection of medical data and reports is a difficult task. So, the dataset used is a combination of five open-source datasets.

## 2. RELATED WORK

On 11th March 2020, The World Health Organization (WHO) declared the virus COVID-19 outbreak as pandemic and since then the virus has spread rapidly in various countries around the world, fatal in many.[12] Symptoms of COVID-19 are typically associated with the symptoms of pneumonia, which can be detected from radiography and imaging tests. Among these two, COVID-19 detection uses image testing in a fast and efficient way when it comes to commercial and wide-scale usage and can therefore be used to control the spread of the virus. Chest X-ray (CXR) and Computed Tomography (CT) are the imaging techniques that play an important role in the diagnosis of COVID-19 disease.

With the technological advancement in the processing of radiography images and image testing (Chest X-ray), more and more machine learning algorithms based on deep learning[13] are being proposed giving promising results in terms of accuracy in detecting COVID-19 from radiography imaging among infected patients. The primary focus is on the CT imaging[14][15][16][17]

Even with the initial release of the proposed open-sourced COVID-Net, many research scholars and institutions face difficulty in accessing public research literature and are unavailable to gather a deeper understanding and extension of these algorithms and models.

However, significant efforts are being made worldwide recently for open access and open source of machine learning models for COVID-19 positive detection from the radiography-driven dataset.[18][19][20] with an exemplary effort being the open-source COVID-19 Image Data Collection, an effort by Cohen et al.16 to build a dataset consisting of COVID-19 cases including severe acute respiratory syndrome (SARS) and the Middle East respiratory syndrome (MERS) cases) with annotated CXR and CT images so that the research community and citizen data scientists can leverage the dataset to explore and build machine algorithms for COVID-19 detection.

A number of research experiments have been conducted in the past 6 months in the area of SARS-COVID19 detection using chest X-ray images following the public release of the proposed COVIDx and COVID-Net dataset.[21][22][23][24][25][26][27][28][29][30][31][32][33]

A detailed study of these proposed research models states that the solution focuses primarily on the in-depth exploration of deep neural networks, specifically deep convolutional neural networks, with results varying depending on the cleanliness of the input data and parameters described in the model for performing the given computer vision task.

## 2.1 Federation Learning

Federated learning is recent emerging research that has been extensively studied in the fields of financial security, artificial intelligence, and robotics.[34][35][36] In this method, instead of loading the training data on the central server, the data is distributed on each block of the server, and only the updated data from each block is added to the central server. Once this data is optimized, the central server can return to the global state of each device and continue to accept the updated data from each block of the server. This technique is called Federated Learning.[37]

With the help of computer vision and the state-of-the-art convolutional neural technique, detection of COVID-19 infection with Chest X-ray images can be successfully conducted on a wide scale. However, the privacy of patients' data must be protected and should not be leaked or shared without any approval. Along with this, the collection of such training data is also a major task.

To a certain extent, this has caused a lack of sufficient data samples when performing deep learning approaches to detect COVID-19. Federated Learning is an available way to address this issue. It can effectively address the issue of data silos and get a shared model without obtaining local data. It is being tested for COVID-19 data training and deploy experiments to verify its effectiveness.

## 2.2 Covid- Net

With the recent advancement in the efficiency and feasibility of machine learning models, studies on imaging analysis and diagnosis on COVID-19 data have increased significantly. Research work is done on the diagnosis of COVID-19 through medical imaging technology. Chest CT scans are used to generate special radiological features and annotations using data augmentation. This is followed by specific machine learning tasks which categorize and identify the medical images generated during Computed tomography scan diagnosis. This significantly reduces the doctor's workload in identifying the pathological characteristics of COVID positive patients from COVID negative patients. Thus, increasing the efficiency of medical institutions and doctors in classifying infected patients. Most of the proposed studies give binary output, that is, "health" and "positive new coronavirus".

COVID - Net is basically a neural network that uses PEPX compression network structure to classify and identify Covid-19 pneumonia CXR images as shown in Fig.1. The network is highly sensitive to the pneumonia symptoms of COVID-19, allowing it to showcase high accuracy during the performance of the network. Based on the advantages of CXR imaging for rapid triage of COVID-19 screening and availability, the network makes predictions through the COVID-Net interpretability method allowing the network to thoroughly analyse the key symptoms of COVID cases and factors related to it.

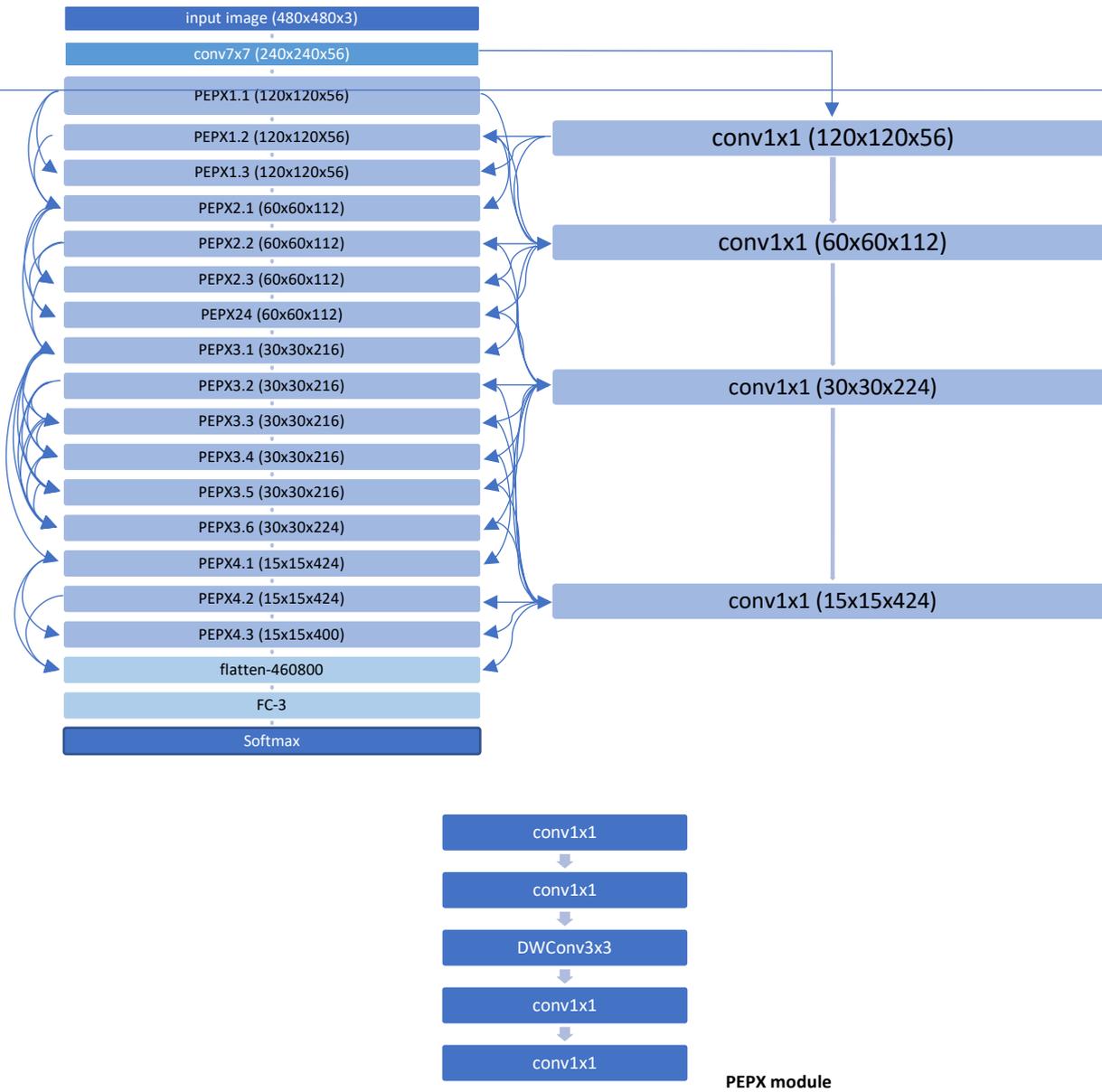

**Figure 1:** COVID-Net architecture. High architectural diversity and selective long-range connectivity can be observed as it is tailored for COVID-19 case detection from CXR images. The heavy use of a projection-expansion-projection design pattern in the COVID-Net architecture can also be observed, which provides enhanced representational capacity while maintaining computational efficiency.

## 3. DATASET

The dataset used to train the proposed model is a combination of two open-source datasets called COVID-19 image data collection and Covid-Net. The COVID-19 image data collection[38] was initially built by collecting medical images from websites and research publications such as Radiopaedia.org, the Italian Society of Medical and Interventional Radiology.[39][40][41][42][43][44]

| Attribute | Description |
| --- | --- |
| Patient ID | Initial identifier |
| Offset | Number of days since start of symptoms or hospitalization for each image. If a report indicates "after a few days", then 5 days is assumed. |
| Sex | Male (M), Female (F), or blank |
| Age | Age of patient in years |
| Finding | Type of pneumonia |
| Survival | Yes (Y) or No (N) |
| View | Posteroanterior (PA), Anteroposterior (AP), AP Supine (APS), or Lateral (L) for X-rays; Axial or Coronal for CT scans |
| Modality | CT, X-ray, or something else |
| Date | Date on which image was acquired |
| Location | Hospital name, city, state, country |
| Filename | Name with extension |
| doi | Digital object identifier (DOI) of the research article |
| url | URL of the paper or website where the image came from |
| License | License of the image such as CC BY-NC-SA. Blank if unknown |
| Clinical notes | Clinical notes about the image and/or the patient |
| Other notes | e.g., credit |

**Table 1:** Descriptions of each attribute of metadata

It also included medical images directly from the PDF using the tool pdf images. The objective of using an imaging tool was to maintain the quality of the images. The COVID-19 image data collection dataset consisted of 148 frontal view Chest X Radiology images.

The Covid-Net dataset is a collection of 13,975 CXR images of around 13,870 infected patients. It is the largest open-access benchmark dataset in terms of images with COVID-19 infected patients.
The dataset also comprises images from leading medical databases including COVID-19 Chest X-ray Dataset Initiative[45]ActualMed COVID-19 Chest X-ray Dataset Initiative[46], established in collaboration with ActualMed RSNA Pneumonia Detection Challenge dataset[47] and COVID-19 radiography database[48]

The combined dataset after filtration consists of 9,438 Chest X-Ray images which are categorized into 2 labels, that is, Covid (Pneumonia) and normal.

(a) 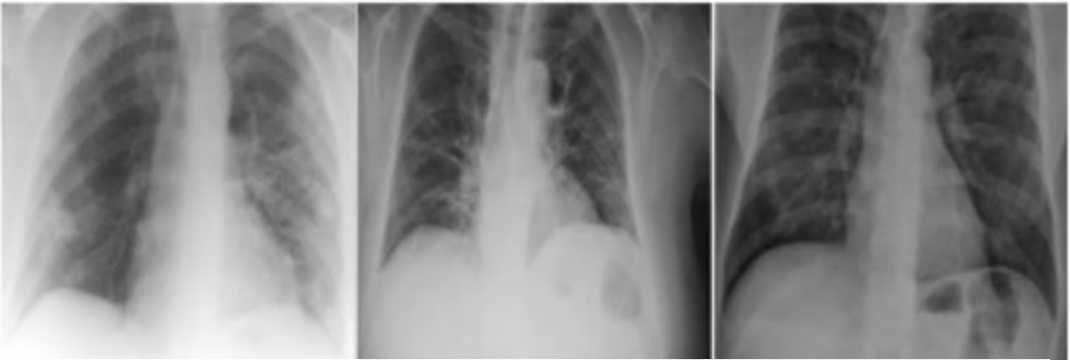

(b) 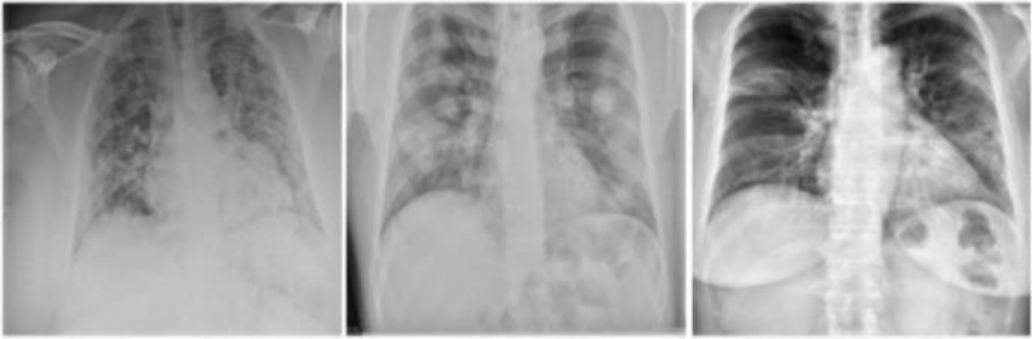

(c) 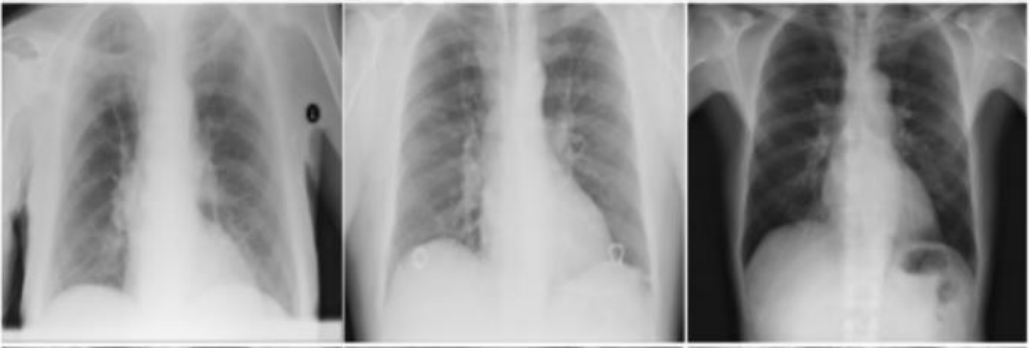

(d) 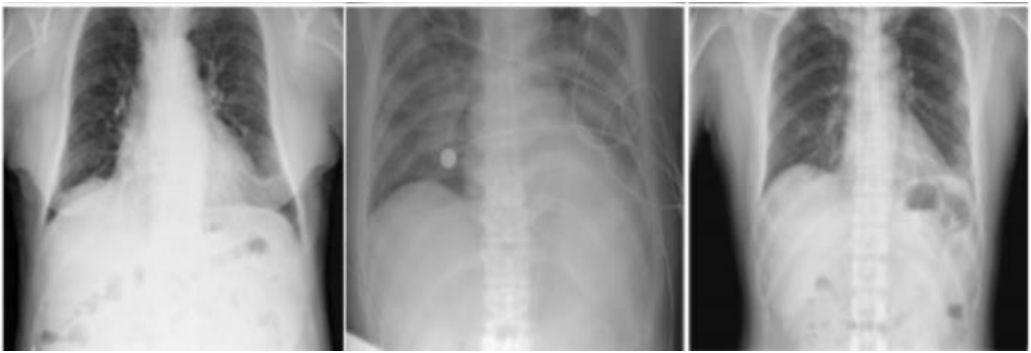

(e) 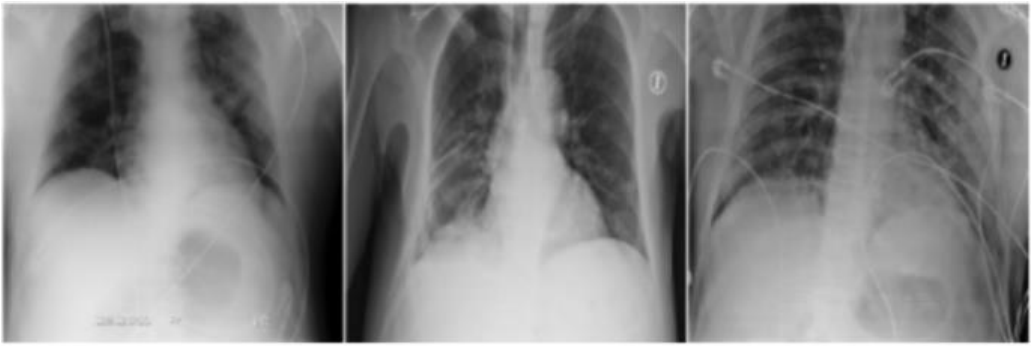

**Figure 2**: Example Chest X-Ray images from the dataset, which comprises of 13,975 Chest X-Ray images across 13,870 patient cases from five open access data repositories: (a) COVID-19 Image Data Collection, (b) COVID-19 Chest X-Ray Dataset Initiative, (c) RSNA Pneumonia Detection challenge dataset, (d) ActualMed COVID-19 Chest X-Ray Dataset Initiative, and (e) COVID-19 radiography database.

Dataset generation is available publicly for open access at https://github.com/lindawangg/COVID-Net.

## 4. DATA PRE-PROCESSING AND AUGMENTATION

During the data-pre-processing step, all the chest CXR images from the image dataset were cropped (top 8% of the image) prior to training of the model to mitigate commonly-found embedded textual information in the CXR images. Further, in order to avoid overfitting and increase the size of the dataset, data augmentation was leveraged with the following properties:

1. Randomly rotate some training images by 30 degrees
2. Randomly Zoom by 20% some training images
3. Randomly shift images horizontally by 10% of the width
4. Randomly shift images vertically by 10% of the height
5. Randomly flip images horizontally.

We perform a grayscale normalization to reduce the effect of illumination's differences

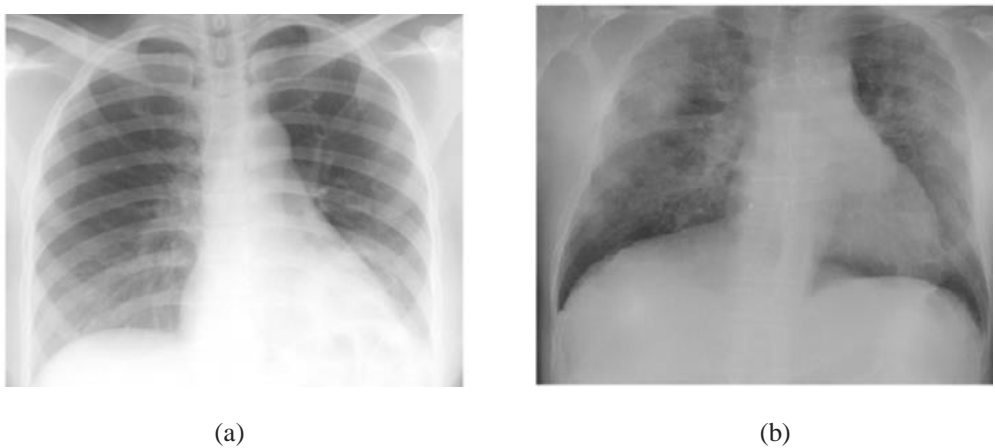

(a)            (b)

**Figure 3:** Example Chest X-Ray images of: (a) non-COVID19 infection, and (b) COVID19 viral infection from the dataset.

## 5. PROPOSED METHOD

In this paper, we have proposed a deep convolutional neural network as shown in fig 4.

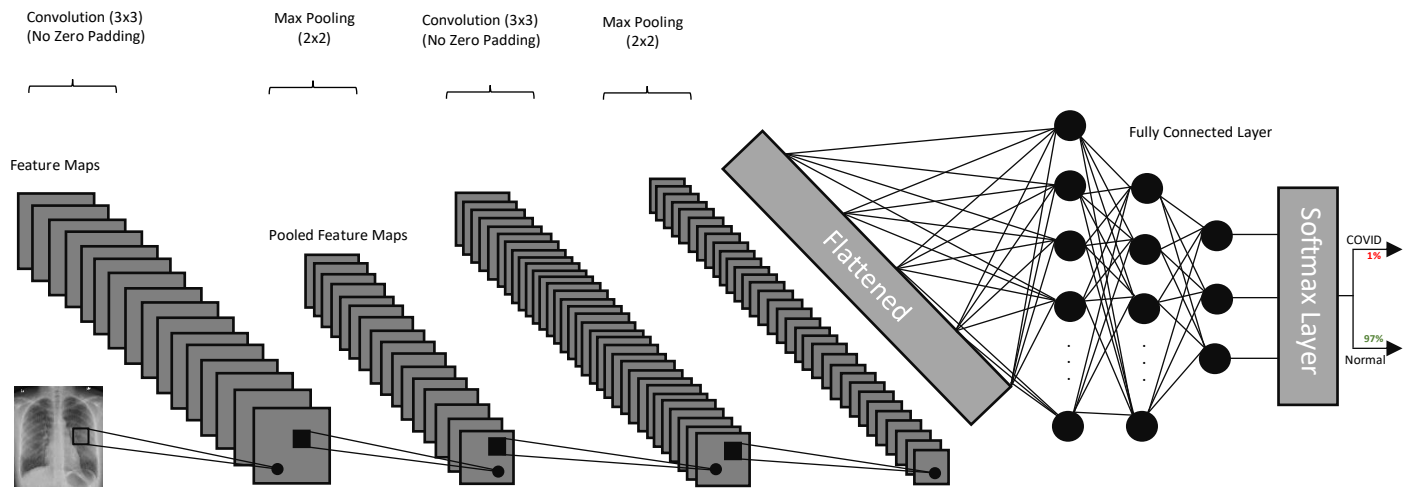

**Figure 4:** Architecture of the proposed deep convolutional neural network

**5.1 Deep Convolutional Neural Networks**

A convolutional neural network is a class of deep neural networks used for image recognition problems and analysis.[49] A subset of the convolutional neural network is the Deep Convolutional Neural network (CNN), which a unique type of deep neural network. The network has performed exceptionally well in many competitions and events themed on Computer Vision and Image analysis. Real-life applications of the deep convolutional neural network include Object detection and segmentation, image classification, Speech recognition, and natural language processing. Deep CNN uses multiple feature extraction at different stages that automatically learns the features represented in the image of the dataset, giving it an edge over basic CNN when it comes to accuracy and efficiency.

Deep learning models are used worldwide for performing commercial tasks such as classification, detection, and segmentation of medical data. These machine learning models are trained on data obtained from medical imaging techniques such as X-ray, Computed Tomography (CT), and Magnetic Resonance Imaging (MRI). Some widely used medical application includes a diagnosis of diseases like a brain tumor, skin cancer, and breast cancer.[50][51][52][53][54][55] With the recent advancement in hardware technology and the availability of exabytes of data, the research on the architecture of deep CNN has accelerated significantly.

For the working of a deep CNN, the input images from the dataset are converted to matrix format first so that the image can be recognized by the computer and then be processed. The images in the matrices are labelled based on their relative difference. The model learns the effects of these differences on the label during the training phase and makes the required predictions on the testing image dataset.

The basic architecture of deep CNN includes three different types of layers, namely convolutional layer, pooling layer, and fully connected layer. The extraction of features from the input images are done during the starting of convolutional and pooling layers. Whereas the process of classification is done in the fully connected layer. Apart from these layers, the architecture of deep CNN also includes 3 basic components, and these are Activation function, Batch Normalization, and Dropout.

The input Chest X-Ray images from dataset has dimensions of 150 x 150 x 1. The optimizer used for the following proposed model is 'Root mean square propagation' (rmsprop) with 'binary cross entropy' as the loss function.

### 5.1.1 Convolutional Layer

Also known as the base layer of CNN, the convolutional layer is responsible for feature extraction and pattern recognition of input images. Images from the training dataset are passed through a filter consisting of feature maps and kernels, which helps the CNN to extract low-level and high-level features.[56] The kernel present in the filter layer is a matrix of size either 3x3 or 5x5, which is transformed with the input features matrix. The output of the convolutional layer can be given as:

$$f_l^k(p,q) = \sum_c \sum_{x,y} i_c(x,y) \cdot e_l^k(u,v) \qquad (1)$$

where, $i_c(x,y)$ is an element of the input image tensor $I_c$, which is element wise multiplied by $e_l^k(u,v)$ index of the $k^{th}$ convolutional kernel $k_l$ of the $l^{th}$ layer.

The stride parameter is the number of steps tuned for shifting over the input matrix (We have set stride parameter to 2 in the proposed model). Since the convolutional layer has the ability to share weights, sliding kernels are used to extract different sets of features with the same set of weights, making deep CNN significantly efficient in terms of accuracy and loss generation.

### 5.1.2 Pooling Layer

After the features are extracted in the convolutional layer, their approximate position relative to others is much more important as compared to their exact location. The pooling layer, also known as a, down-sampling layer, is then used to aggregate similar information in the neighbourhood of the feature layer and outputs the dominant response within this layer

$$\mathbf{Z}_l^k = g_P(\mathbf{F}_l^k) \qquad (2)$$

Equation (2) shows the pooling operation in which $\mathbf{Z}_l^k$ represents the pooled feature-map of $l^{th}$ layer for $k^{th}$ input feature-map $\mathbf{F}_l^k$, whereas $g_p(.)$ defines the type of pooling operation. The use of pooling operation helps to extract a combination of features, which are invariant to $x_i$

The use of the pooling layer in the deep CNN is to extract a combination of features, which are invariant to translational shifts and small distortions.[57] A reduction in the dimension of feature-map to invariant feature set can help the prediction model to reduce the possibility of overfitting and regulate the complexity of the neural network layers. Different types of pooling formulations such as max, average, L2, overlapping, spatial pyramid pooling, etc. are used in CNN.

We have used max pooling technique in the proposed model with a dimension of 2 x 2.[58] There is also a global average pooling layer that is only used before the fully connected layer, reducing data to a single dimension. It is connected to the fully connected layer after the global average pooling layer.

### 5.1.3 Fully connected layer

The fully connected layer is used at the end of the deep CNN network for classification and is the most important layer of CNN. The function of the fully connected layer is a global operation, like a multi-perceptron, unlike the pooling and the convolutional layer which operates locally. The input is the features extracted from different stages of the network and is then compared and analysed with the outputs of all the preceding layers. Mathematical computation of these two activation functions is as follows:

$$\text{ReLU}(x) = \begin{cases} 0, & x < 0 \\ x, & x \geq 0 \end{cases} \qquad (2)$$

$$\text{Soft max}(x_i) = \frac{e^{x_i}}{\sum_{y=1}^{m} e^{x_y}} \qquad (3)$$

Where $x_i$ and m represent input data and the number of classes, respectively.

Neurons in a fully connected layer have full connections to all activation functions in the previous layer. The most common activation function used on a fully connected layer is the Rectified Linear Unit (ReLU) activation function. Whereas the Softmax activation function is used to predict output images in the last layer of the fully connected layer.

### 5.1.4  Activation function

The role of the activation function is to transform the weighted sum input of one node for a layer and use it for the activation of the node for the particular input. The activation function serves as a decision function, helping in the learning of feature patterns.

The activation function for a convolved feature-map is defined in equation

$$T_l^k = g_a(F_l^k) \qquad (4)$$

In the above equation, $F_l^k$ is the output of a convolution which is assigned to activation function $g_a(.)$ that adds non-linearity and returns transformed output $T_l^k$ for $l^{th}$ layer.

Some commonly used activation functions are sigmoid, tranh, softmax, maxout, SWISH, ReLU, and Leaky ReLu. MISH activation functions, proposed recently, are also used due to their better performance as compared to ReLu.[59] In this research, we will be using ReLu and its variants function for their ability to overcome the vanishing gradient problem, which arises frequently in CNN models. Sigmoid function will be used in the final layer of the fully connected layer for binary classification.

### 5.1.5  Batch Normalization

Batch normalization is a layer that allows every layer of the network to do learning more independently. It is used to normalize the output of the previous layers, which helps in addressing the issues related to the internal covariance shift within the feature-maps. The internal covariance shift is a change in the distribution of hidden units' values, which slows down the convergence and requires careful initialization of parameters.

Batch normalization for transformed feature map $F_l^k$ is shown in equation (5)

$$N_l^k = \frac{F_l^k - \mu_B}{\sqrt{\sigma_B^2 + \varepsilon}} \qquad (5)$$

In equation (5), $N_l^k$ represents normalized feature-map, $F_l^k$ is the input feature-map, $\mu_B$ and $\sigma_B^2$ 2depict mean and variance of feature-map for mini batch, respectively. In order to avoid division by zero, is added for numerical stability.

Batch normalization is used in CNN models to prevent and regulate overfitting, making the model much more efficient. The layer is implemented to standardize the inputs and outputs and can be used at several points in between the CNN layers. Batch normalization layer is usually placed just after the sequential model and after the convolutional and pooling layers.

### 5.1.6  Dropout

Dropout is used to add regularization within the deep CNN network, which proves the generalization by skipping some units or connections with a certain probability at random.

This removal of some random connections or units generates several thinned network architectures, and out of these, one network with small weights is selected. And this selected architecture is then taken as an approximation of all of the proposed networks. The dropout parameter set for the proposed model is 0.2.

### 5.1.7  Flatten Layer

Flatten Layer is used to convert the pooled feature map into a 1-dimensional array as an input to the next layer in the form of a single long feature vector. This is connected to the deep neural network, called as a fully connected layer. In the proposed model, the flatten layer has converted the output convolutional layer into a single 1-dimensional array of size 128 units.

| Layer (type) | Output Shape | Param # |
|---|---|---|
| conv2d_56 (Conv2D) | (None, 150, 150, 32) | 320 |
| batch_normalization_56 | (Batc (None, 150, 150, 32) | 128 |
| max_pooling2d_44 | (MaxPooling (None, 75, 75, 32) | 0 |
| conv2d_57 (Conv2D) | (None, 75, 75, 64) | 18496 |
| dropout_45 (Dropout) | (None, 75, 75, 64) | 0 |
| batch_normalization_57 | (Batc (None, 75, 75, 64) | 256 |
| max_pooling2d_45 | (MaxPooling (None, 38, 38, 64) | 0 |
| conv2d_58 (Conv2D) | (None, 38, 38, 64) | 36928 |
| batch_normalization_58 | (Batc (None, 38, 38, 64) | 256 |
| max_pooling2d_46 | (MaxPooling (None, 19, 19, 64) | 0 |
| conv2d_59 (Conv2D) | (None, 19, 19, 128) | 73856 |
| dropout_46 (Dropout) | (None, 19, 19, 128) | 0 |
| batch_normalization_59 | (Batc (None, 19, 19, 128) | 512 |
| max_pooling2d_47 | (MaxPooling (None, 10, 10, 128) | 0 |
| conv2d_60 (Conv2D) | (None, 10, 10, 256) | 295168 |
| dropout_47 (Dropout) | (None, 10, 10, 256) | 0 |
| batch_normalization_60 | (Batc (None, 10, 10, 256) | 1024 |
| max_pooling2d_48 | (MaxPooling (None, 5, 5, 256) | 0 |
| flatten_12 (Flatten) | (None, 6400) | 0 |
| dense_23 (Dense) | (None, 128) | 819328 |
| dropout_48 (Dropout) | (None, 128) | 0 |
| dense_24 (Dense) | (None, 1) | 129 |

**Table 2:** Layer wise architecture of the proposed deep convolutional neural network

## 6. EXPERIMENTATIONS AND RESULTS

During the experiment, the accuracy and loss of both the training and validation set were recorded for 12 consecutive epochs. The value of each category for every epoch run is shown in the Table 3. It was observed that the model had an accuracy of 96.82% on the training set and an accuracy of 92.63% on the validation set as shown in Table 3(a). The loss and validation loss generated during the experiment is shown in Table 3(b). The loss generated during the training and validation set are 0.0971 and 0.293 respectively.

| Epochs | Training Accuracy | Validation Accuracy |
|---|---|---|
| 1 | 0.8357 | 0.5 |
| 2 | 0.8982 | 0.5 |
| 3 | 0.9135 | 0.5625 |
| 4 | 0.9296 | 0.5 |
| 5 | 0.931 | 0.5 |
| 6 | 0.9513 | 0.5625 |
| 7 | 0.9538 | 0.6875 |
| 8 | 0.9641 | 0.57 |
| 9 | 0.963 | 0.7567 |
| 10 | 0.9622 | 0.6 |
| 11 | 0.9641 | 0.8722 |
| 12 | 0.9682 | 0.9263 |

(a)

| Epochs | Loss | Validation Loss |
|---|---|---|
| 1 | 0.5957 | 38.5047 |
| 2 | 0.2812 | 30.2517 |
| 3 | 0.2294 | 19.2671 |
| 4 | 0.2118 | 28.6478 |
| 5 | 0.1951 | 1.6 |
| 6 | 0.1388 | 2.5989 |
| 7 | 0.1344 | 23.3912 |
| 8 | 0.1109 | 1.3816 |
| 9 | 0.1078 | 3.0119 |
| 10 | 0.122 | 0.4287 |
| 11 | 0.0965 | 1.097 |
| 12 | 0.0971 | 0.293 |

(b)

**Table 3:** Accuracy and loss obtained on the training and validation set respectively for 12 epochs.

The Fig 5 represents the graph of Epochs versus Accuracy. Accuracies generated by the model on both the sets – training set and validation set – are marked on the graph during the 12 epochs time period.

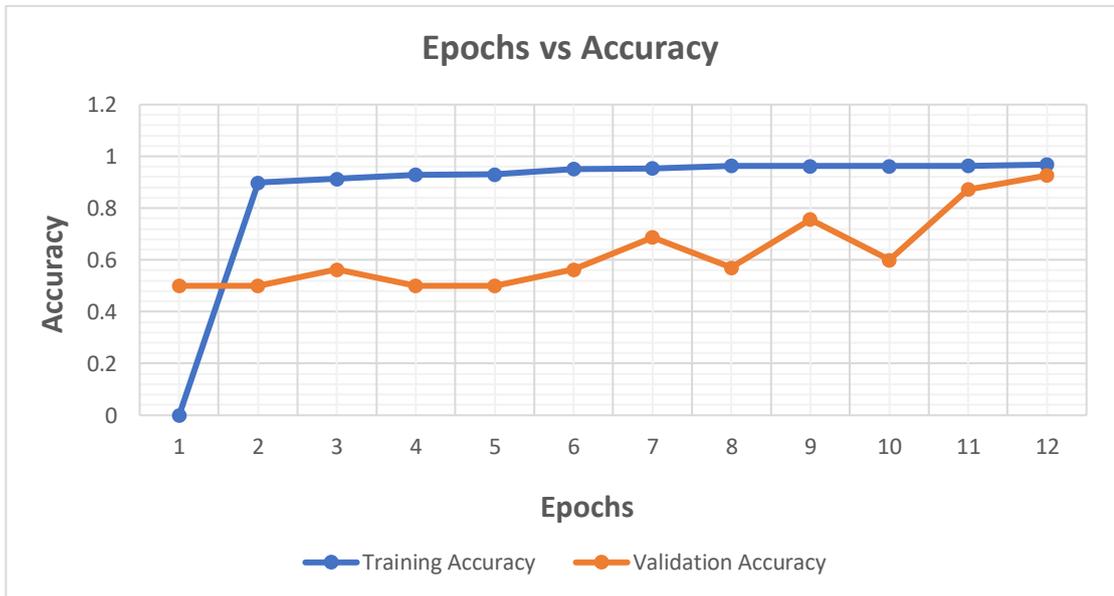

**Figure 5:** Accuracy obtained during 12 epochs

The Fig 6 represents the graph of Epochs versus Loss. Losses generated by the model on both the sets – training set and validation set – are marked on the graph during the 12 epochs time period.

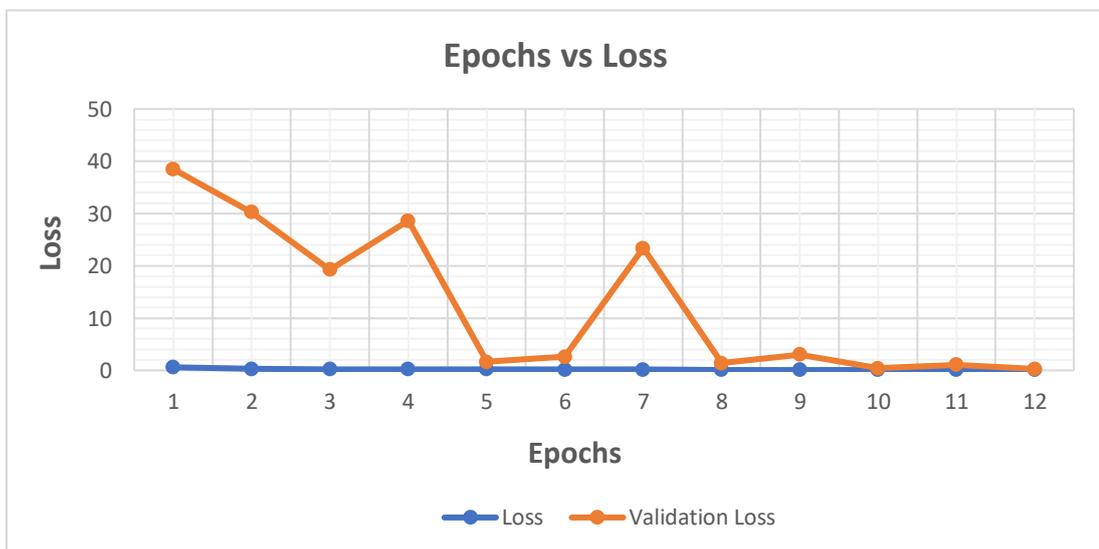

**Figure 6:** Graph plotted for Loss generated during 12 epochs

10 criteria were used for the performances of the proposed deep convolutional network as shown in Table 4. These are Sensitivity, Specificity, Precision, Negative Predictive Values, False Positive Rate, False Discovery Rate, False Negative Rate, Accuracy, F1 Score, and Matthews Correlation Coefficient where TP, FP, TN and FN represent the number of True Positive, False Positive, True Negative and False Negative, respectively.

For the given dataset, TP is the proportion of positive (COVID-19) that are correctly labelled as COVID-19 by the model; FP is the proportion of negative (normal) that are mislabelled as positive (COVID-19); TN is the proportion of negative (normal) that are correctly labelled as normal and FN is the proportion of positive (COVID-19) that are mislabelled as negative (normal) by the model.

| Measure | Value | Derivations |
| --- | --- | --- |
| Sensitivity | 0.9187 | TPR = TP / (TP + FN) |
| Specificity | 0.9378 | SPC = TN / (FP + TN) |
| Precision | 0.9576 | PPV = TP / (TP + FP) |
| Negative Predictive Value | 0.8829 | NPV = TN / (TN + FN) |
| False Positive Rate | 0.0622 | FPR = FP / (FP + TN) |
| False Discovery Rate | 0.0424 | FDR = FP / (FP +TP) |
| False Negative Rate | 0.0813 | FNR = FN / (FN + TP) |
| Accuracy | 0.9263 | ACC = (TP + TN) / (P + N) |
| F1 Score | 0.9378 | F1 = 2TP / (2TP + FP +FN) |
| Matthews Correlation Coefficient | 0.8485 | TP*TN - FP*FN / sqrt((TP+FP) *(TP+FN) *(TN+FP) *(TN+FN)) |

**Table 4:** Performance Matrix

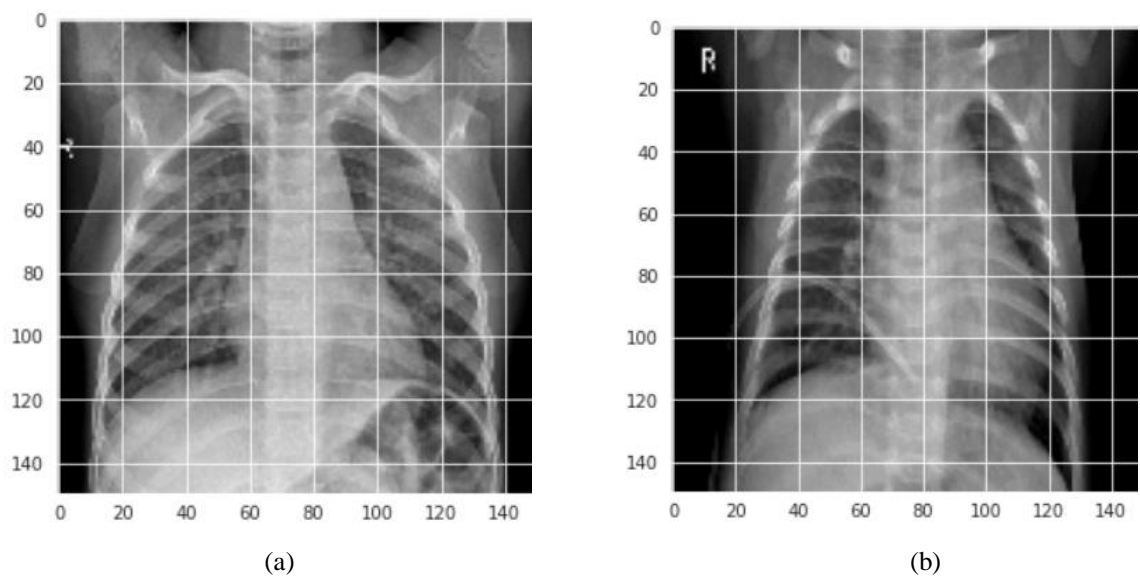

(a)      (b)

**Figure 7:** Normal and Covid19 Chest X-Ray output images

Figure 7 shows the output generated by the proposed deep convolutional neural network. Fig 7 (a) and Fog 7 (b) shows the correctly predicted normal chest X-ray (non-covid) image and covid positive chest X-Ray image, respectively.

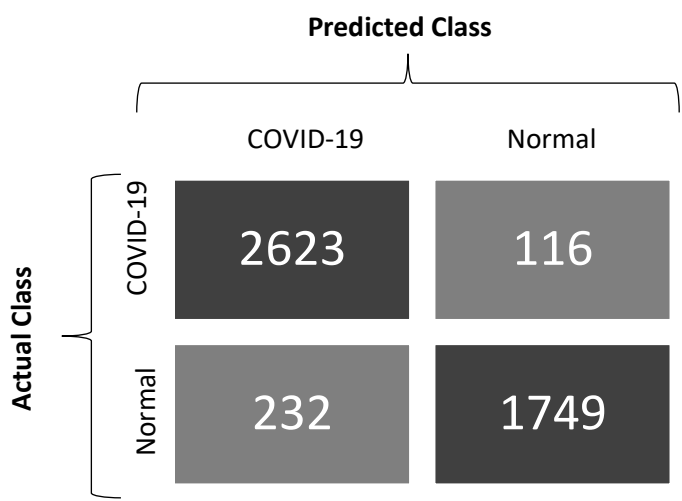

**Figure 8:** Confusion Matrix

The final accuracy and loss generated by the proposed model is compared with four different models which are being used currently for commercial and research purposes. The four models are -

1. COVID-Net: A neural network specifically proposed to identify COVID-19 pneumonia CXR images utilizes PEPX to compress the network structure while preserving the network's performance to a large extent. At the same time, it has high sensitivity to the pneumonia characteristics of COVID-19.

2. ResNet18: Used as residual network, an identity mapping layer is combined with the basic deep convolutional neural network to increase the depth of the fully connected layers. This helps to prevent the accuracy of the model to fall as it reduces the chances of overfitting.

3. ResNet: It is based on residual neural network using split-transform-merge strategy to convert single-core convolution into multi-core convolution, but the topology is the same as ResNet18.

4. MobileNet-v2: Different from residual network, MobileNet-v2 is a lightweight model. In contrast to ResNet18 and ResNet which first uses convolutional kernel to compress and extract features and then expand, MobileNet-v2 comparatively expands and extracts more features and then compresses.

| Model | Accuracy (Testing Set) | Loss Convergence (testing Set) |
|---|---|---|
| Proposed Model | 92.62±0.015% | 0.293 |
| COVID-Net | 89.17±0.015% | 0.261 |
| ResNet18 | 91.26±0.014% | 0.112 |
| ResNet | 90.37±0.015% | 0.157 |
| MobileNet-v2 | 86.83±0.017% | 0.287 |

**Table 5:** Proposed model's accuracy and loss comparison with COVID-Net, ResNet18, ResNeXt and MobileNet-v2 respectively.

On analysing the results from the Table 5, MobileNet-V2 has a larger accuracy gap while ResNet18 has a smaller accuracy gap. If considering the number of parameters at the same time, MobileNet and ResNet18 have a higher performance. MobileNet has the fewest parameters and the lowest accuracy. ResNet18 has the second fewest parameters and the highest accuracy while COVID19 has the highest parameters but less accuracy as compared to ResNet18, ResNet and the proposed model.

## CONCLUSION

In this work, we have proposed a deep convolutional neural network designed specifically for the detection of COVID-19 cases by implementing computer vision and image analysis on Chest X-Ray images gathered from five open access data repositories. The dataset comprised of 9472 chest X-Ray images from more than 13,870 patients. We conducted experiments on COVID19 identification and compared it with four models: COVID-Net, ResNet18, ResNet and MobileNet-v2, and comparison experiment between the accuracy and loss generated on the validation set by each model. The experimental results show that the proposed model had the best performance accuracy on the validation set. Further, we investigated and applied different model parameters in order to gain deeper insights on the Chest X-Ray features critical for classifying Covid and non-Covid patients which can aid clinicians in improved screening as well as improve trust and transparency.

However, the proposed model, by no means is a commercially-ready solution. The hope is that the promising results obtained from this work can be leveraged and further be improvised by both researchers and data scientists to accelerate the development of highly accurate yet practical deep learning solutions for detecting COVID-19 from Chest X-Ray images and accelerate of those in need.

## ACKNOWLEDGEMENT

This work was supported by Birla Institute of Technology and Science, Pilani, Dubai Campus, Dubai, UAE. Special thanks to Professor Vilas H Gaidhane, Head of Department, Electrical & Electronics Engineering, BITS Pilani Dubai Campus.